\documentclass[cits]{JINST}

\usepackage{graphicx}
\usepackage{multirow}
\usepackage{booktabs}

\title{Delayed avalanches in Multi-Pixel Photon Counters}

\author{K. Boone$^{a,!}$, Y. Iwai$^{b}$, F. Reti\`ere$^a$\thanks{Corresponding author.}, and C. Rethmeier$^{a,c}$\\
\llap{$^a$}TRIUMF, 4004 Wesbrook Mall, Vancouver, BC, V6T2A3, Canada\\
\llap{$^b$}Hamamatsu Corporation, 2875 Moorpark Ave Ste 200, San Jose, CA 95128\\
\llap{$^c$}University of Victoria, Victoria, BC , Canada\\
\llap{$^!$}Now at University of California Berkeley, Berkeley, CA, USA\\
E-mail: \email{fretiere@triumf.ca}}

\abstract{
Hamamatsu Photonics introduced a new generation of their Multi-Pixel Photon Counters in 2013 with significantly reduced after-pulsing rate. In this paper, we investigate the causes of after-pulsing by testing pre-2013 and post-2013 devices using laser light ranging from 405 to 820nm. Doing so we investigate the possibility that afterpulsing is also due to optical photons produced in the avalanche rather than to impurities trapping charged carriers produced in the avalanches and releasing them at a later time. For pre-2013 devices, we observe avalanches delayed by ns to several 100~ns at 637, 777nm and 820 nm demonstrating that holes created in the zero field region of the silicon bulk can diffuse back to the high field region triggering delayed avalanches. On the other hand post-2013 exhibit no delayed avalanches beyond 100~ns at 777nm. We also confirm that post-2013 devices exhibit about 25 times lower after-pulsing. Taken together, our measurements show that the absorption of photons from the avalanche in the bulk of the silicon and the subsequent hole diffusion back to the junction was a significant source of after-pulse for the pre-2013 devices. Hamamatsu appears to have fixed this problem in 2013 following the preliminary release of our results. We also show that even at short wavelength the timing distribution exhibit tails in the sub-nanosecond range that may impair the MPPC timing performances.
}


\begin{document}

\section{Motivation}

Pixelated Geiger-mode avalanche Photodiodes (PPDs) are increasingly replacing conventional Photo-Multiplier Tubes (PMTs) in numerous applications including particle physics experiments, nuclear physics experiments and medical imaging (see for example ~\cite{Renker, GaruttiSiPMHEP, SiPMNuclMed, SiPMPET}). For example, the T2K near detector that was completed in 2010 used about 50,000 Multi-Pixel Photon Counters (MPPCs) from Hamamatsu Photonics~\cite{T2KNIMPaper}. In addition to being compact and insensitive to magnetic fields, PPDs and MPPCs in particular achieve excellent photodetection efficiency of up to 35\% and large gain making it possible to to identify individual photo-electrons (for example see ~\cite{Vacheret2011}). For most applications until 2013, the main limitation of MPPCs was not the dark noise rate but the probability of generating additional avalanches following a primary avalanche created either by a photo-electron or a thermal carrier (i.e. dark noise). Indeed the average number of correlated avalanches created by the original avalanche approached or exceeded 1 when biasing the MPPC at 0.5 to 1V over the Hamamatsu recommended operating voltage (corresponding to a gain of 7.5 10$^5$). In such conditions thermally generated carriers could produce large pulses corresponding to more than 10 photo-electrons often mimicking real physical pulses. While operating above the recommended operating voltage may not appear desirable, it has been shown that highest photodetection efficiency~\cite{Vacheret2011}, and best timing resolution~\cite{Ronzhin2010, Collazuol2007} are achieved above this limit. In fact, in applications that typically detect a lot of photons, such as Positron Emission Tomography, the MPPCs will most certainly be operated well above the recommended operating voltage because 10-20 photo-electron pulses from dark noise are not a significant nuisance. However, even in this case, correlated avalanches may eventually limit the energy resolution. Hence, it is desirable to reduce the correlated avalanche rate in order to achieve optimum performances.

In this paper, we investigate the concept of delayed cross-talk introduced in \cite{Acerbi2015}. We question the assumption that after-pulsing is due to charge carriers (hole and electrons) that are produced during the avalanches getting trapped on impurities and then released at a latter time when the MPPC voltage is sufficiently high to generate subsequent avalanches as discussed for example in~\cite{Cova1991, Jensen2006}. We investigate the possibility that after-pulsing is due to optical photons created in the avalanche that are believed to be responsible for the cross-talk process. Experimentally, cross-talk and after-pulsing are two well separated processes; cross-talk avalanches occur within less than one nanosecond of their parent while afterpulsing avalanches occur several ns to 100 ns after their parent. Cross-talk is believed to originate from optical photons produced in the parent avalanche, subsequently absorbed in the high field region of a neighboring pixel, hence triggering new avalanches~\cite{Buzhan2009}. Furthermore, in the same paper Buzhan et al. show that optical photons may not only generate prompt avalanches if they have a direct line of sight with neighboring pixel high field regions, but also generate delayed avalanches if carriers are allowed to drift from the zero-field silicon bulk to the high field region. Hence, afterpulsing, as defined experimentally as correlated avalanches occurring after their parents, may also be due to optical photons generated in the avalanche, the delay being due to the charge carrier diffusion time in the silicon bulk.

The wavelength spectrum of the optical photons has been measured in~\cite{Mirzoyan2009} for Hamamatsu MPPCs. At the recommended operating voltage, each avalanche produces about 9 photons having an energy sufficient for creating an electron-hole pair. Such photons are mostly at wavelength longer than 600 nm; hence their absorption length in silicon exceeds a few microns, and they will be mostly absorbed in the silicon bulk. The question is whether or not the carriers, holes in the case of the MPPC, produced in the silicon bulk can diffuse to the high field region and produce an avalanche. In this paper, we address this question by measuring the timing distribution of avalanches produced by photons with 5 different wavelengths, 405, 467, 637, 777 and 820 nm. We expect to measure delayed avalanches with the longer wavelengths if holes are able to drift back to the high field region. Our data also allow us to measure the single photon timing resolution, hence verifying earlier measurements reported in~\cite{Ronzhin2010}. The experimental results for Hamamatsu pre 2013 devices are reported in section 2.  In section 3, we interpret the data inferring the hole lifetime and structure of the high field region. In section 4, we perform similar tests in order to compare the pre and post 2013 devices. Conclusions are drawn in section 5.

\section{Pre-2013 MPPC timing response to several wavelengths}

The experimental setup consisted of a laser, the MPPC, an amplifier and an oscilloscope. Two different laser setups were used: a Hamamatsu C10196 pulser coupled with Hamamatsu M10306 laser heads for wavelengths of 637 nm and below and a Hamamatsu PLP-01 pulser with an LDH-082 laser head for 820 nm light. Both lasers were run at frequencies of 10kHz. The light was transported through an optical fiber that was setup to illuminate the MPPCs uniformly. The amount of light per pulse was adjusted to yield less than 1 photon per pulse on average using a variable digital attenuator. A different laser head produced each wavelength. The relative delay between the laser electrical pulse and the light flash coming from each laser head was measured with a Hamamatsu photo-multiplier tube (PMT). Unfortunately, the PMT had no sensitivity at 820 nm, and the delay could not be calibrated out in this case.

The output of the MPPC was connected to a custom amplifier providing a gain of 100 and a rise time of 4 ns. The MPPC was biased using a Keithley 6487 picoammeter/voltage source. The output of the amplifier was connected to a Tektronix MSO 5204 oscilloscope which triggered on a reference pulse from the laser and sampled at 10GS/s for 1$\mu$s. The MPPC pulse corresponding to the light flash occurred 200~ns from the start of the waveform, hence allowing the detection of delayed avalanches up to 800~ns after the prompt pulse. The entire setup was kept inside of a Cincinnati Sub-Zero temperature chamber (model MCBH-1.2-.33-.33-H/AC) in order to maintain a constant temperature between -60$^\circ$C and 20$^\circ$C to within 0.1$^\circ$C.

Three different MPPCs were used. The initial analysis was performed using the MPPCs designed for the T2K experiment. These devices have 50$\mu$m$ \times $50$\mu$m pixels and an active area of 1.3mm $\times$ 1.3mm. The properties of these MPPCs can be found in \cite{Kyoto2010} and~\cite{Vacheret2011}. In order to further understand the processes being examined, two more devices were tested: a Hamamatsu S10362-11-100P and a Hamamatsu S10362-11-100P LDC. These devices have identical geometries with 1mm $\times$ 1mm active faces and 100$\mu$m$ \times $100$\mu$m pixels. However, the LDC device comes from a batch that was produced with reduced impurity level. It exhibits a significantly lower dark noise rate that may translate into reduced after-pulsing due to trapping but a longer charge carrier life time. According to the specification sheets from Hamamatsu, the S10362-11-100P has a dark noise count of 543kHz and the S10362-11-100P LDC has a dark noise count of 154kHz at 25$^\circ$C at nominal over-voltage and a 0.5 photoelectron threshold.

The waveforms are analysed in software. A pulse finder algorithm identifies avalanches using the derivative of the waveform. Then, the times of each avalanche are calculated by using a digital constant fraction discriminator algorithm;  the program defines the avalanche time as the time at which the pulse reaches one fifth of its peak value. The analysis subtracts the baseline before the avalanche in order to ensure that the timing is not affected by earlier avalanches. The number of pixels avalanching is determined from the charge in each pulse. Events with only one pulse within the 1 $\mu$s window are selected. This eliminates events with after-pulsing altogether and causes dark noise pulses to be uniformly distributed in time. The time of each pulse is then added to a histogram. The noise level is dependent on the probability of detecting a photon because in order to accept a dark noise pulse no photo-electron triggered avalanche must occur within the 1 $\mu$s window. The dark noise contribution is subtracted out from the distribution by measuring the noise level in the first 200 ns of the recorded waveforms where no photo-electron triggered avalanches occur. The histogram is then normalized so that its integral is equal to one. In other word, we measure the probability distribution function of detecting a photo-electron at time t within an 800 ns window following the light flash.

The dark noise rates were measured by sampling for 200 nanoseconds before the start of the main peak. Due to the fact that the counts are normalized, the dark noise rate varies for the different
wavelengths. At -60$^\circ$C, the dark noise rate was found to be between $5\times 10^{-6}$ and $2 \times 10^{-5}$ normalized counts/ns for the different
wavelengths. The tails seen at 820 nm and 637 nm are above this level by several orders of magnitude for several hundreds of nanoseconds so these tails cannot be due to dark noise. At 20$^\circ$C, the dark noise levels are between $5 \times 10^{-4}$ and $8 \times 10^{-4}$ normalized counts/ns. The tail approaches these levels after approximately 100 nanoseconds but there is still a significant amount of photons in this tail at longer wavelengths.

\begin{figure}[ht]
  \begin{minipage}[b]{0.5\linewidth}
    \centering
    \includegraphics[scale=0.4]{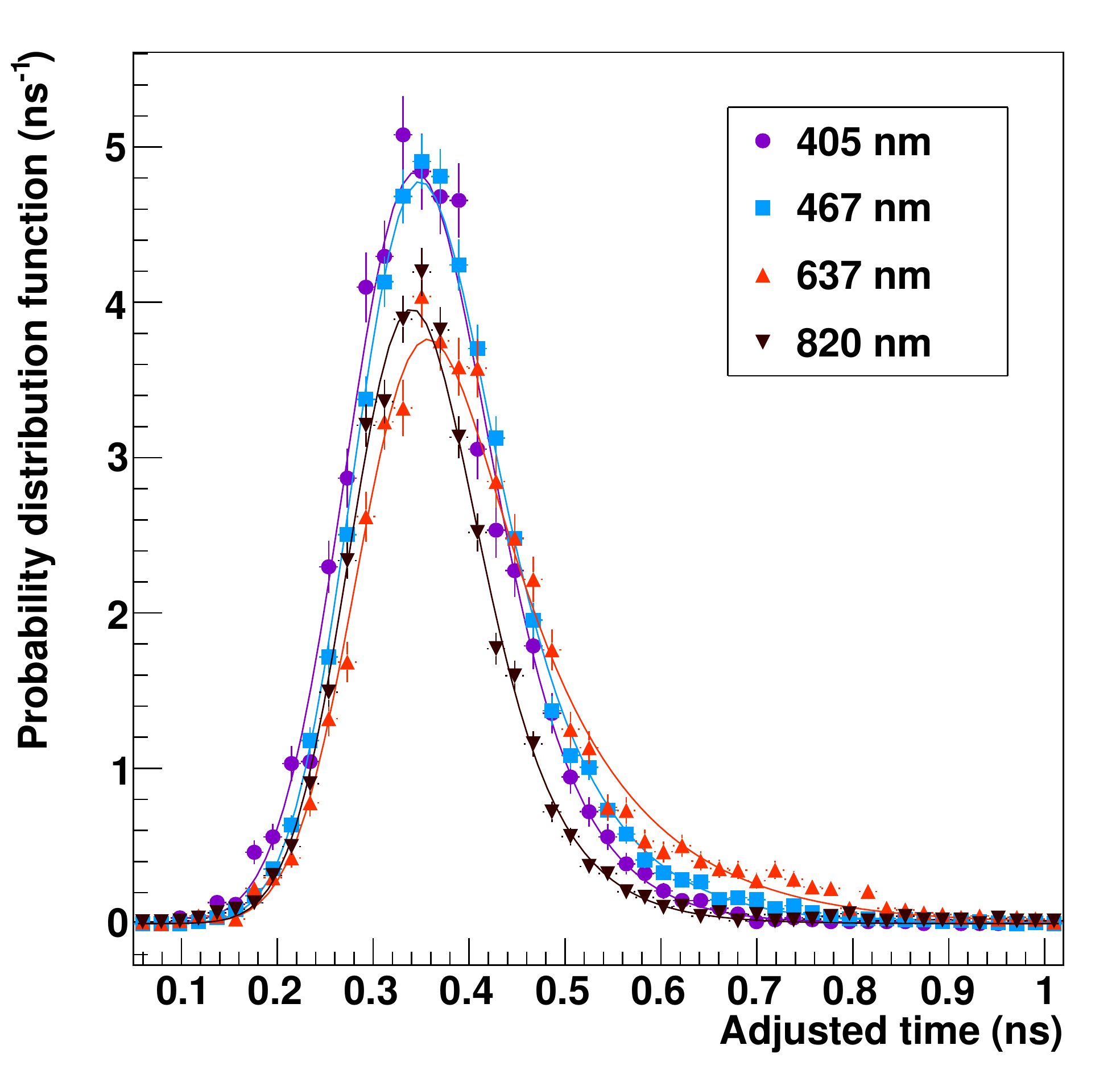}
    \caption{Main peak of the timing distribution for the Hamamatsu T2K MPPC at -60$^\circ$C and an over-voltage of 2.24V. The data are fitted by the convolution of Gaussian and exponential functions. }
    \label{fig:t2k-60Cmain}
  \end{minipage}
  \hspace{0.5cm}
  \begin{minipage}[b]{0.5\linewidth}
    \centering
    \includegraphics[scale=0.4]{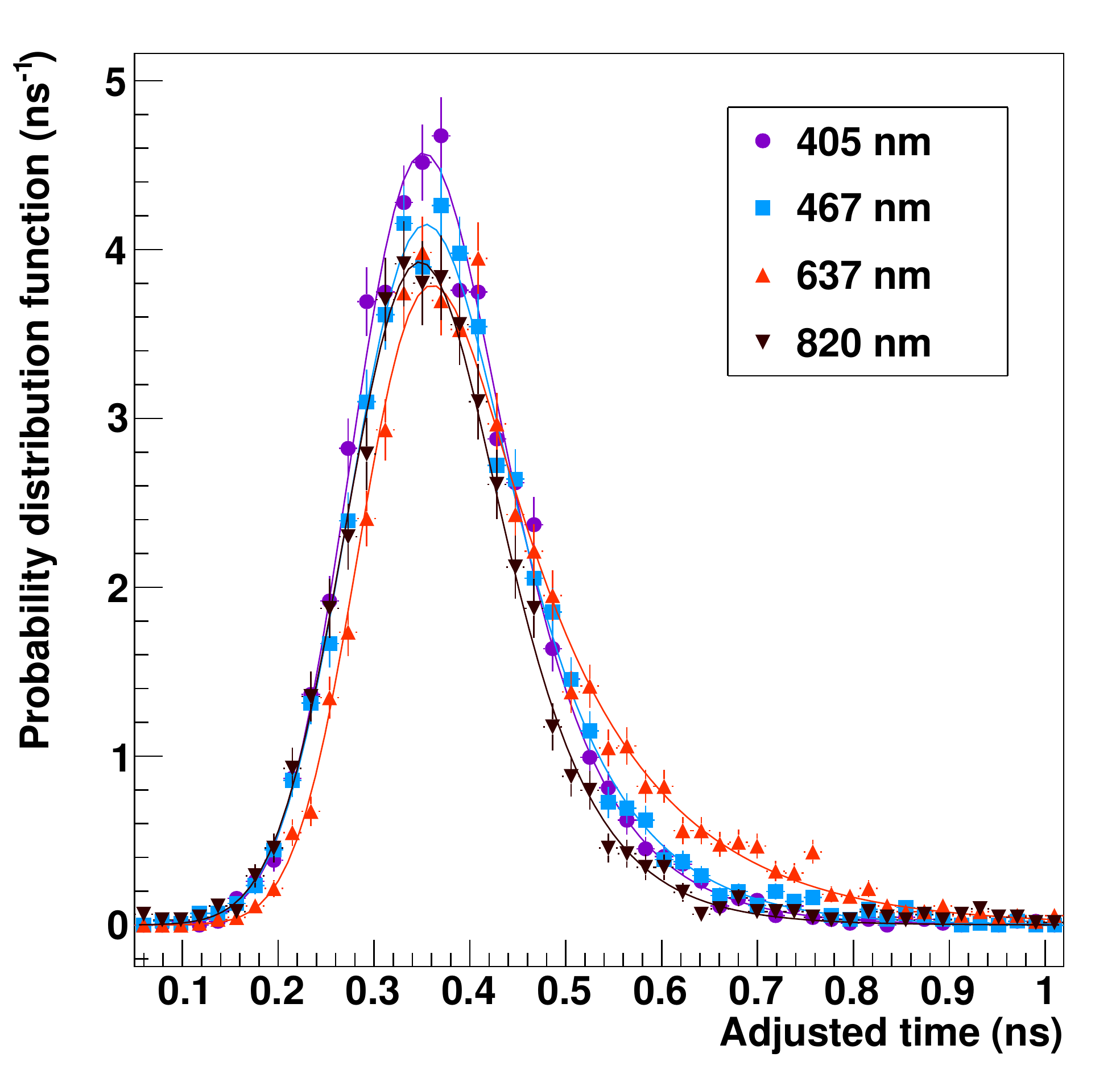}
    \caption{Main peak of the timing distribution for the Hamamatsu T2K MPPC at 20$^\circ$C and an over-voltage of 2.10V. The data are fitted by the convolution of Gaussian and exponential functions}
    \label{fig:t2k20Cmain}
  \end{minipage}
\end{figure}

The probability distribution functions are shown in Figures \ref{fig:t2k-60Cmain} and \ref{fig:t2k20Cmain} for -60$^\circ$C and 20$^\circ$C respectively zooming on the first nanosecond. After aligning the relative time of the laser head using the PMT calibration data, the effective time of the light flash was adjusted in the analysis to occur at t=0.3 ns rather than 0ns so that the data could be shown with a logarithm x axis. The data at 820nm wavelength were adjusted to coincide with the other wavelengths because no PMT data was available. The time of the light flash was obtained by fitting the prompt peak by the convolution of Gaussian and Exponential functions, with the Gaussian $\mu$ parameter corresponding to the average laser flash time. The data were then shifted so that the laser flash time occurred at t=0.3 ns. Overall, the convolution of Gaussian and Exponential functions fit the data very well within the narrow 0-1 ns.

\begin{figure}[ht]
  \begin{minipage}[b]{0.5\linewidth}
    \centering
    \includegraphics[scale=0.4]{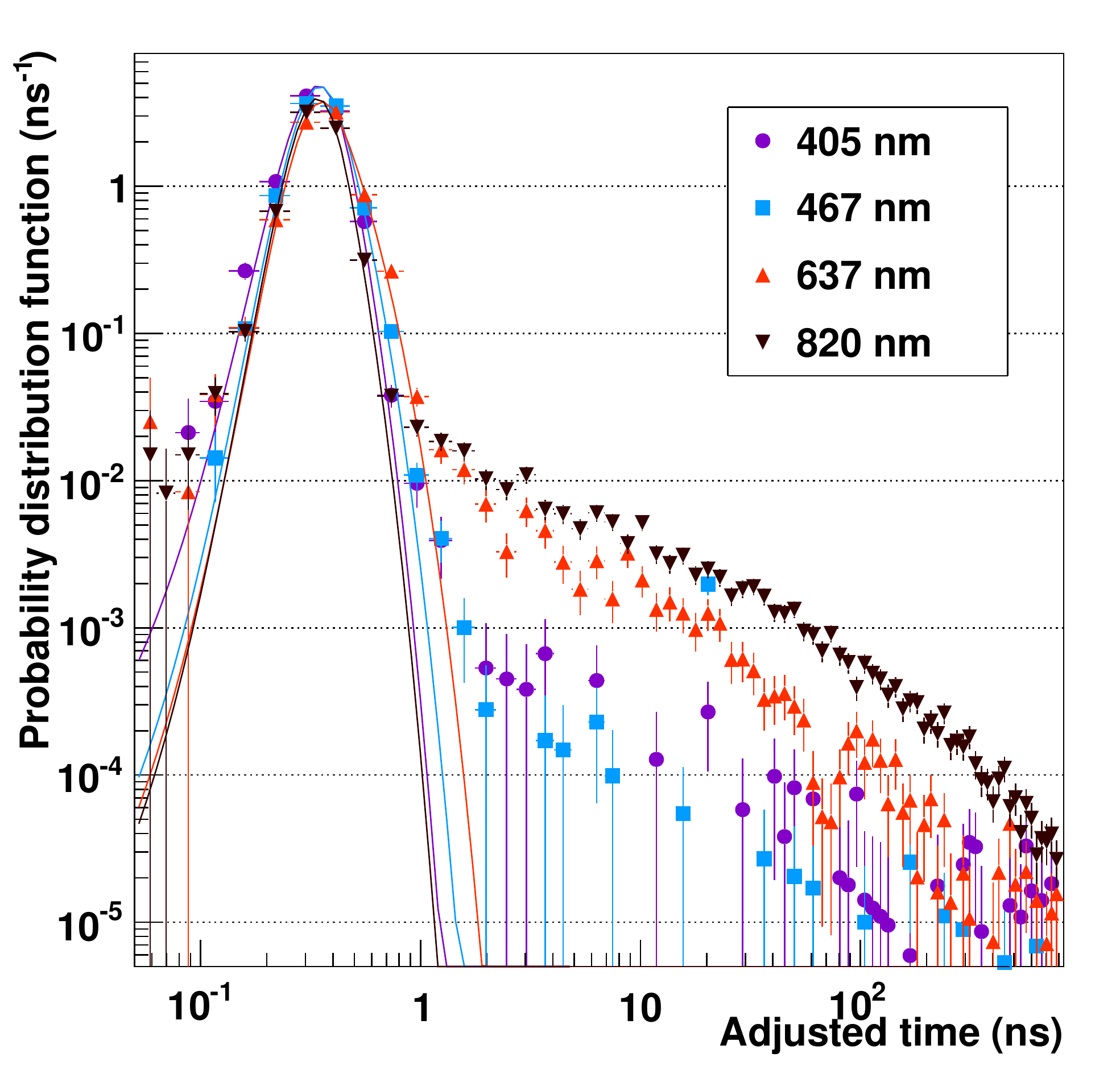}
    \caption{Timing distribution for the Hamamatsu T2K MPPC at -60$^\circ$C and an over-voltage of 1.9V. The position of the prompt peak is arbitrary. The lines show the convolution of Gaussian and Exponential functions obtained by fitting the same data between 0 and 1 ns.}
    \label{fig:t2k-60Call}
  \end{minipage}
  \hspace{0.5cm}
  \begin{minipage}[b]{0.5\linewidth}
    \centering
    \includegraphics[scale=0.4]{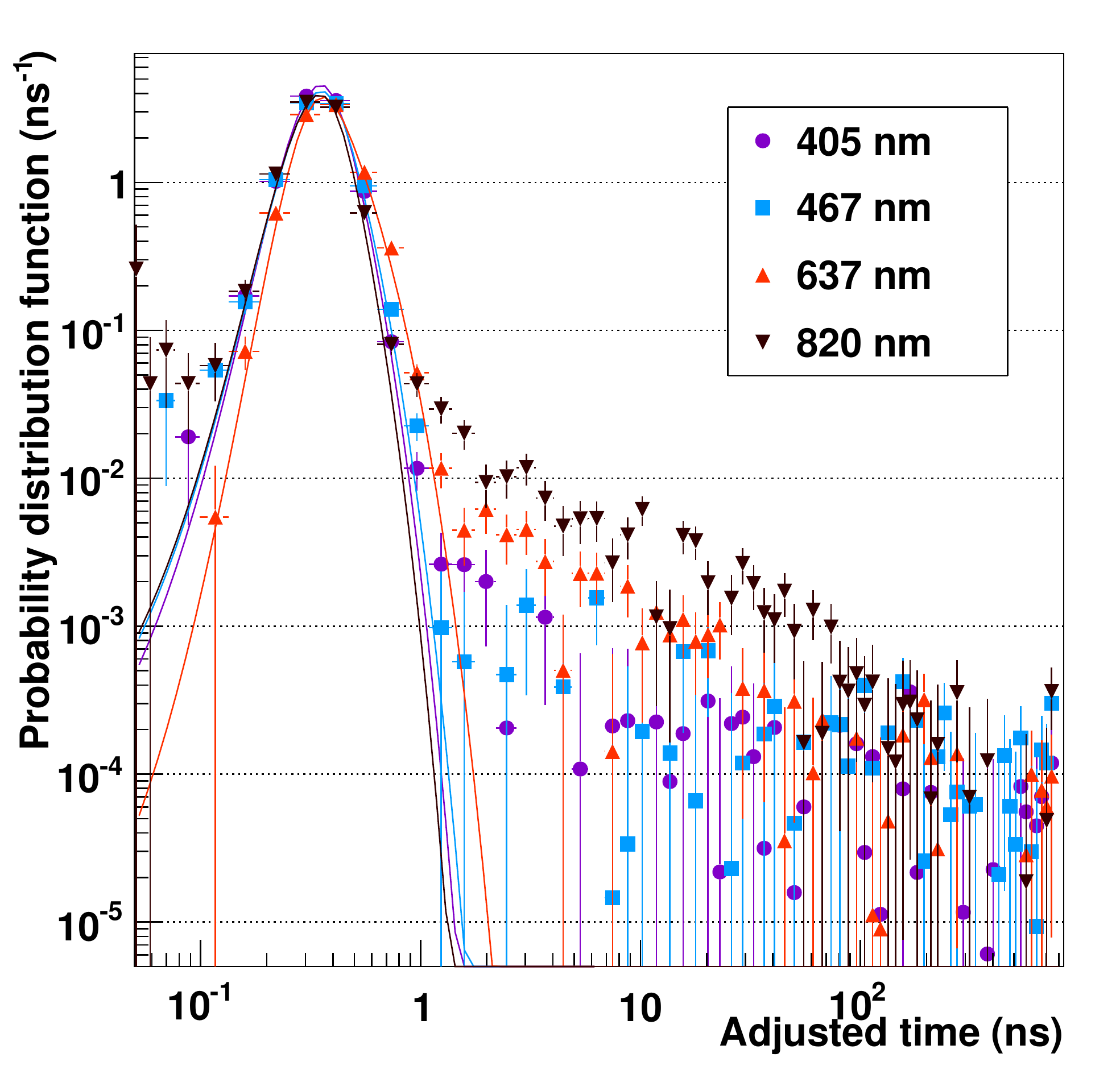}
    \caption{Timing distribution for the Hamamatsu T2K MPPC at 20$^\circ$C and an over-voltage of 2.10V. The lines show the convolution of Gaussian and Exponential functions obtained by fitting the same data between 0 and 1 ns.}
    \label{fig:t2k20Call}
  \end{minipage}
\end{figure}

The full 800 ns window is shown in Figures \ref{fig:t2k-60Call} and \ref{fig:t2k20Call}. The prompt peak is clearly followed by a long tail at 637 and 820 nm at both -60$^\circ$C and 20$^\circ$C with the convolution of Gaussian and Exponential functions failing to reproduce the data beyond 1 ns or so. However the error bars beyond 10 ns at 20$^\circ$C because the dark noise dominates and yield to very large corrections. Hence it is difficult to study delayed avalanche phenomena at room temperature. Fortunately the parameters driving the delayed avalanches, the hole mobility and corresponding diffusion constant, and the hole life time both vary weakly with temperature. Hence the results obtained at -60$^\circ$C can in principle be easily scaled up to room temperature.

\begin{figure}[!h]
  \begin{minipage}[b]{0.5\linewidth}
\vspace{0pt}
    \centering
    \includegraphics[scale=0.4]{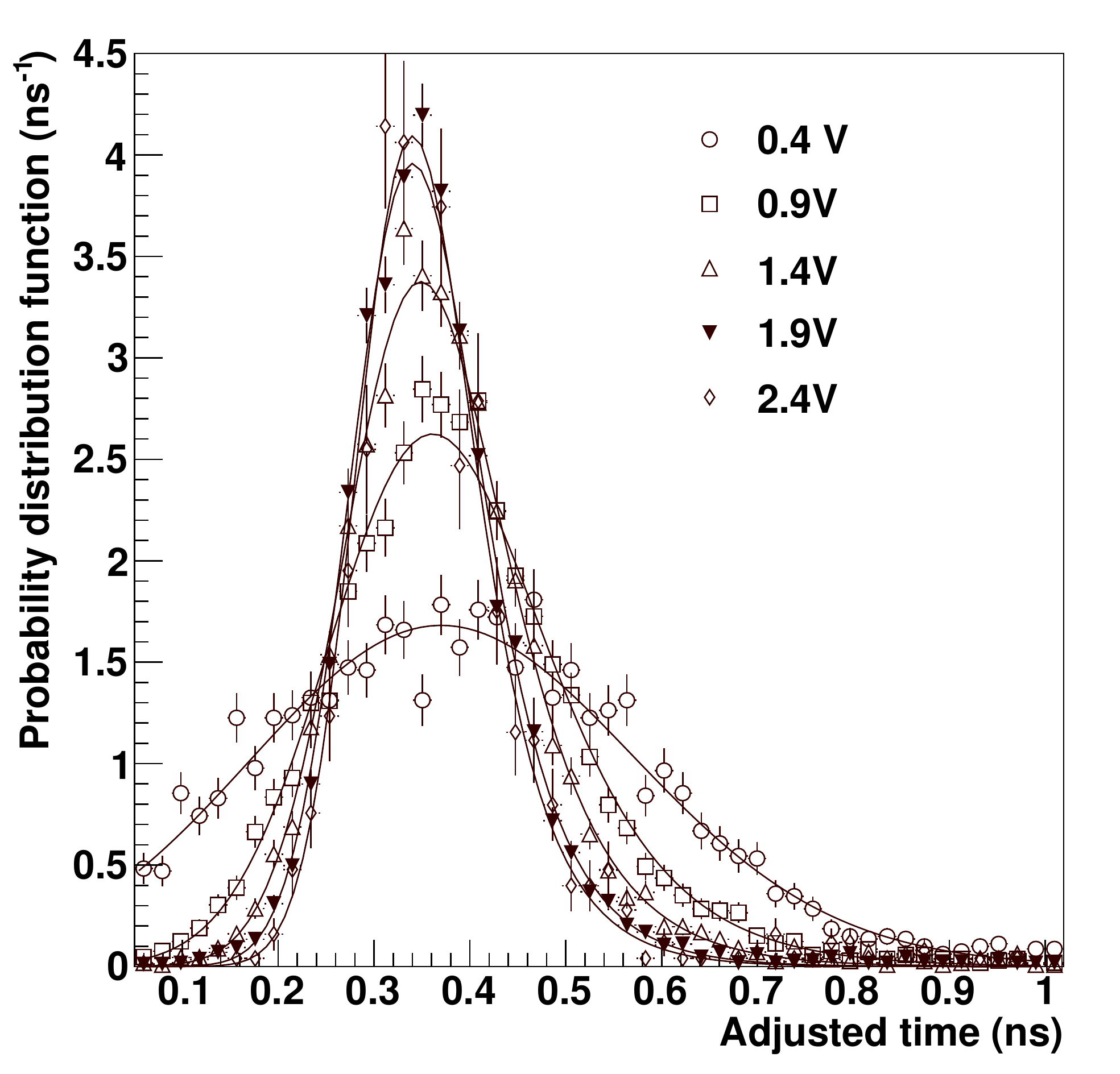}
    \caption{Timing distributions for the Hamamatsu T2K MPPC at -60$^\circ$C as a function of over-voltage.The lines are obtained by fitting the data to the convolution of Gaussian and Exponential functions.}
    \label{fig:t2k-60CDVmain}
  \end{minipage}
  \hspace{0.5cm}
  \begin{minipage}[b]{0.5\linewidth}
\vspace{0pt}
    \centering
    \includegraphics[scale=0.4]{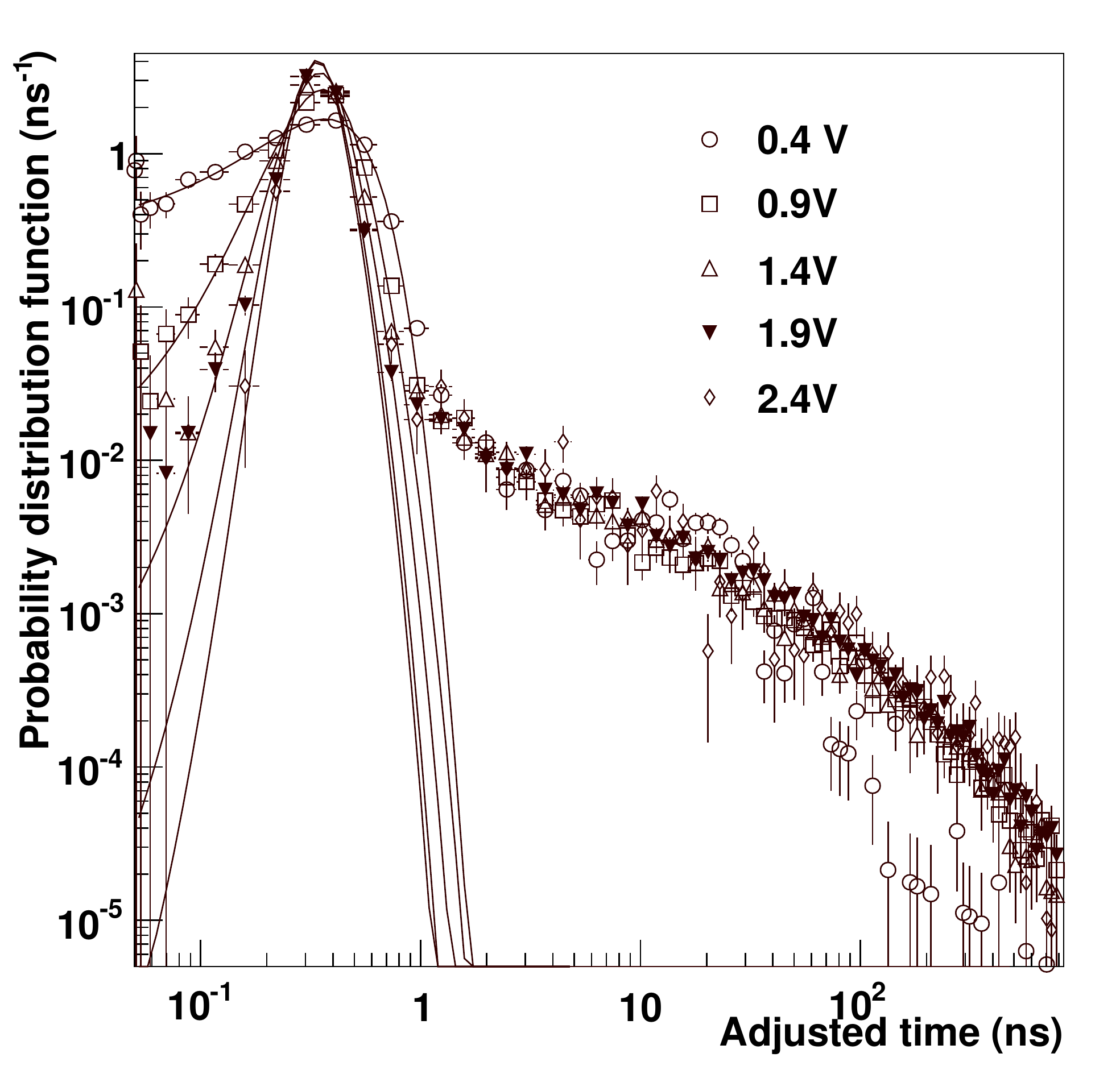}
    \caption{Timing distributions for the Hamamatsu T2K MPPC at -60$^\circ$C as a function of over-voltage. The lines show the convolution of Gaussian and Exponential functions obtained by fitting the same data between 0 an 1 ns.}
    \label{fig:t2k-60DVall}
  \end{minipage}
\end{figure}

\begin{figure}[!t]
  \begin{minipage}[b]{0.3\linewidth}
    \centering
    \includegraphics[scale=0.25]{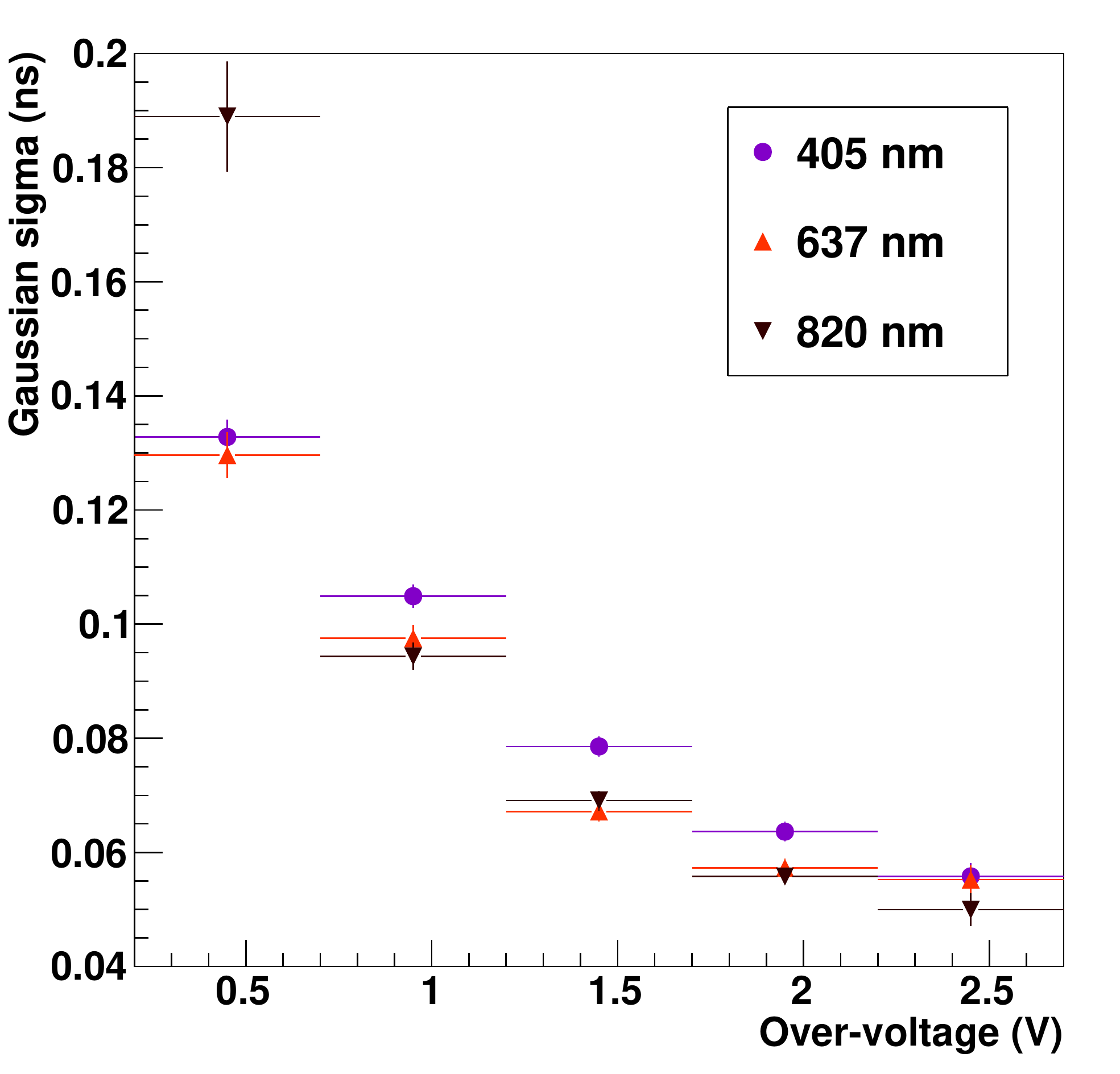}
    \caption{Gaussian sigma as a function of over-voltage for 3 different wavelengths measured at  -60$^\circ$C}
    \label{fig:sigDV}
  \end{minipage}
  \hspace{0.5cm}
  \begin{minipage}[b]{0.3\linewidth}
    \centering
    \includegraphics[scale=0.25]{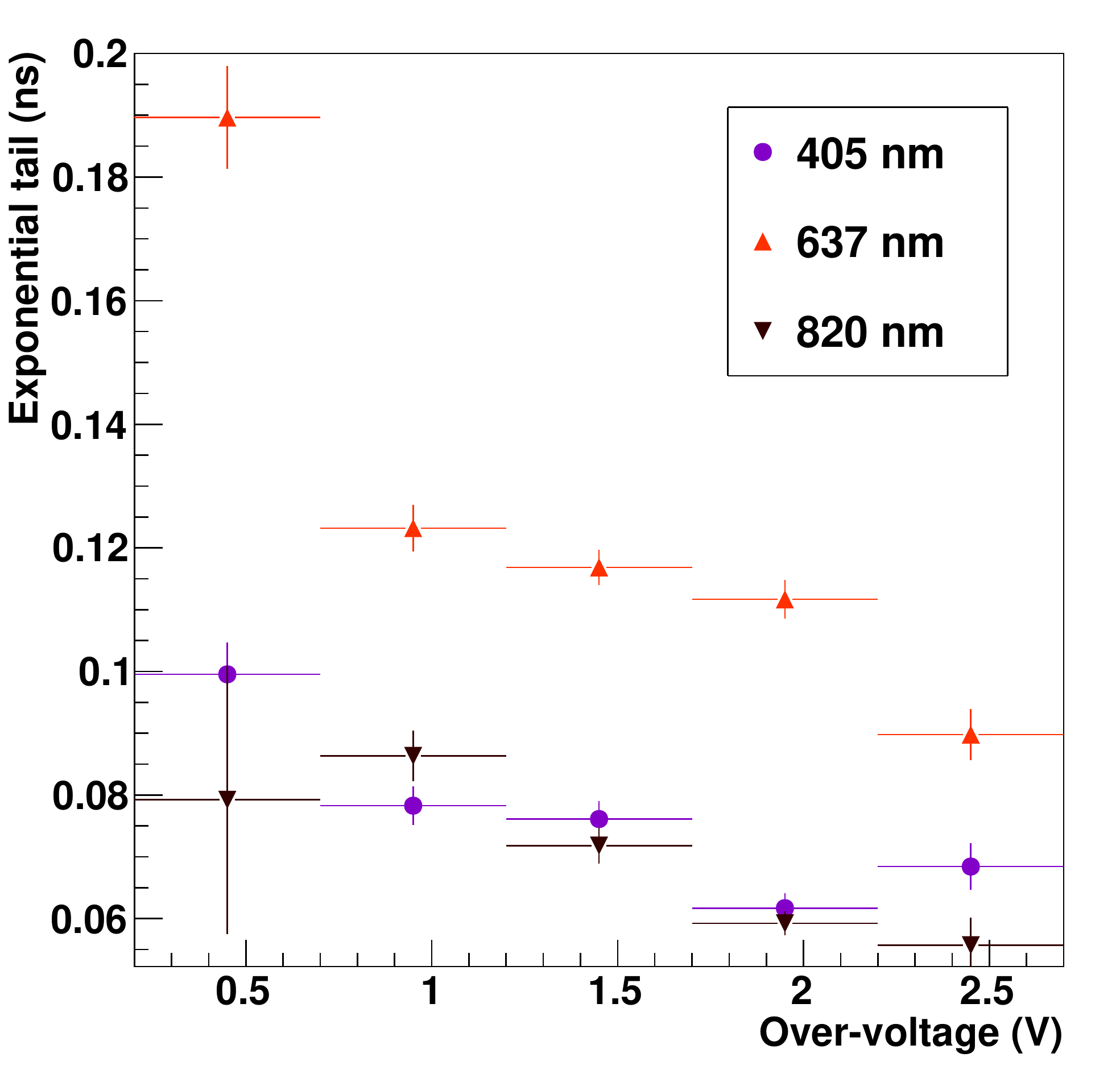}
    \caption{Exponential tail as a function of over-voltage for 3 different wavelengths measured at  -60$^\circ$C}
    \label{fig:tauDV}
  \end{minipage}
  \hspace{0.5cm}
  \begin{minipage}[b]{0.3\linewidth}
    \centering
    \includegraphics[scale=0.25]{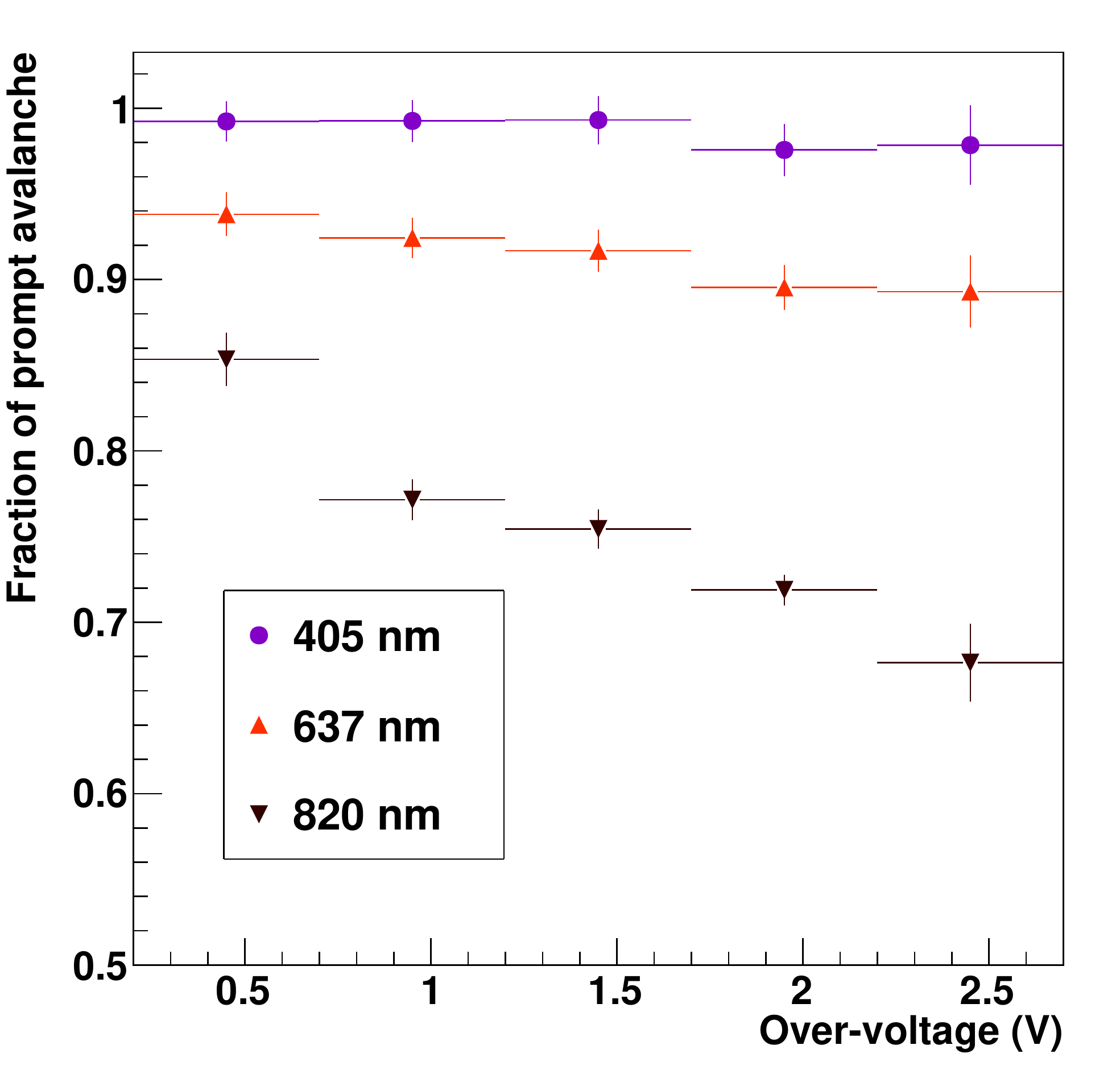}
    \caption{Fraction of prompt light as a function of over-voltage for 3 different wavelengths measured at  -60$^\circ$C}
    \label{fig:fracDV}
  \end{minipage}

\end{figure}

The over-voltage dependence of the probability distribution functions is shown in Figures \ref{fig:t2k-60CDVmain} and \ref{fig:t2k-60DVall}. With the possible exception of the lowest over-voltage point the data are independent of over-voltage beyond 1 ns. On the other hand the shape of the prompt distribution vary with over-voltage. The convolution of Gaussian and Exponential functions fit the data well at all over-voltages. Figures \ref{fig:sigDV}, \ref{fig:tauDV} and \ref{fig:fracDV} show the over-voltage dependence of the Gaussian $\sigma$, the Exponential decay time constant and the fraction of prompt light respectively. To first order, i.e. ignoring the exponential tail, the Gaussian $\sigma$ is the single photon timing resolution. As shown previously~\cite{Ronzhin2010} it decreases with increasing over-voltage reaching 50 ps at the highest over-voltage. The exponential tail also drops with increasing over-voltage. This tail may be explained by arguing that some photons are absorbed in zones having a low enough electric field that the carriers take around 100 ps to travel to the multiplication region. This tail is most prominent at 637 nm, while all the other wavelengths exhibit smaller tails. The fraction of prompt light behaves as expected recalling that all the late light is due to avalanches triggered by holes while the prompt light stems mostly from electron triggered avalanches. As the over-voltage increases the probability that electrons trigger avalanches rapidly approaches 1 due to electron high impact ionization probability. On the other hand, the probability that holes trigger avalanches increases more or less linearly with over-voltage hence increasing the probability of late avalanches while the early (electron triggered) avalanche probability is essentially constant.

\begin{figure}[ht]
  \begin{minipage}[b]{0.5\linewidth}
    \centering
    \includegraphics[scale=0.4, angle=90]{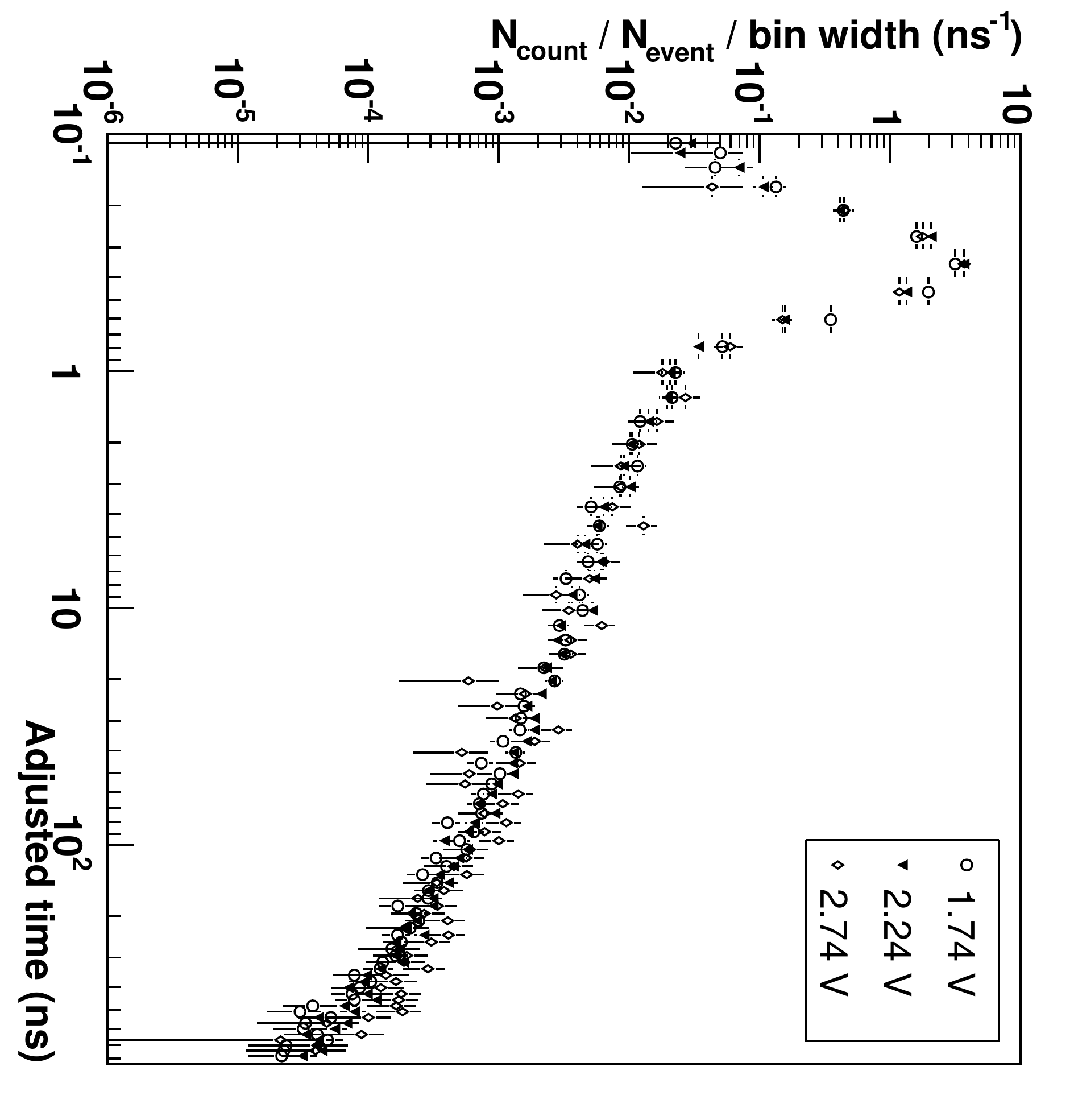}
    \caption{Timing distribution for the Hamamatsu T2K MPPC for different over-voltages using 820nm light at -60$^\circ$C}
    \label{fig:t2kneg60Covervoltage}
  \end{minipage}
  \hspace{0.5cm}
  \begin{minipage}[b]{0.5\linewidth}
    \centering
    \includegraphics[scale=0.4, angle=90]{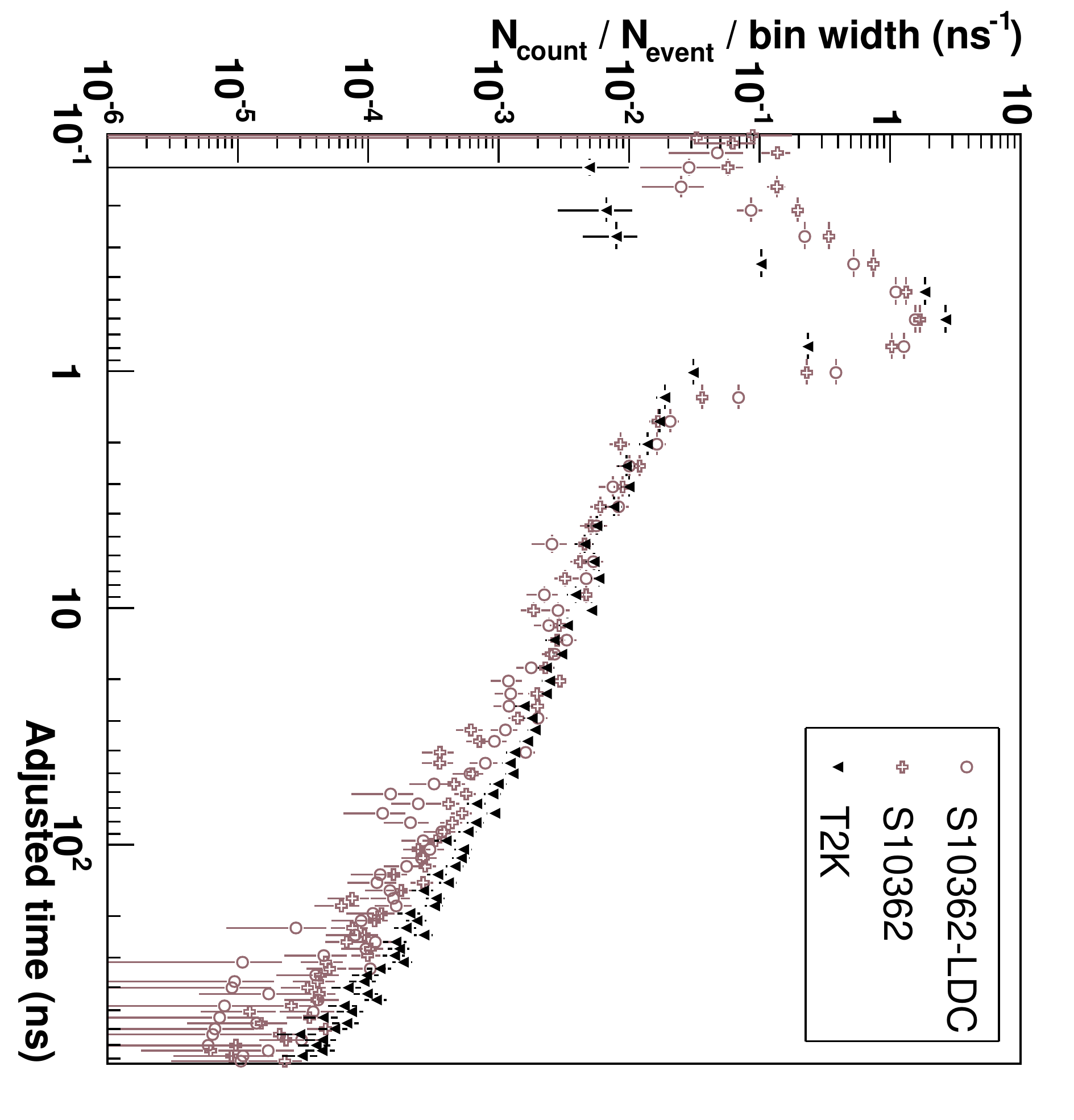}
    \caption{Timing distribution for the different devices using 820nm light at -60$^\circ$C}
    \label{fig:devicescomparison}
  \end{minipage}
\end{figure}

A comparison of the distributions for the three different devices is shown in Figure \ref{fig:devicescomparison}. Both the S10362 device and the S10362-LDC
device have nearly identical distributions which implies that the purity of the silicon that drive the dark noise rate does not affect the hole diffusion pattern. One could have expected the hole lifetime to increase, hence increasing the fraction of late avalanches, but it is not observed. There are significant differences in the timing distribution between the T2K and S10362 devices, which are likely explainable by the different pixel size. In this paper, we will focus on investigating the T2K MPPC, which has been characterized in depth in~\cite{Vacheret2011}.

\begin{figure}[ht]
\end{figure}

\section{Interpretation}

Figure \ref{fig:t2k-60Call} shows that the timing distribution has a significant tail at longer wavelengths. After-pulsing does not play any role in these timing distributions since they are for a single photoelectron. These data indicate that there are several
orders of magnitude difference between the tails using light at wavelengths of 405 nm and 820 nm. If these tails were due to trapping of the primary holes or electrons then there would not be a significant dependence on wavelength. The dependence on wavelength suggests that holes being created in the
silicon bulk and are diffusing back to the multiplication region, triggering delayed avalanches.

A simulation was created in order to determine the timing distribution that would be produced using this hypothesis. The program simulates the effects of a single photon arriving perpendicular to the surface of the MPPC using Monte Carlo techniques. If the photon produces a photoelectron, then the effects of drift and diffusion of the carrier are simulated until it either triggers an avalanche, hits the detector boundaries or is reabsorbed. The simulation takes into account dark noise but ignores the effects of after-pulsing and cross-talk since they do not apply when filtering for single photoelectron events. The objective of this simulation is to model the tail of the distribution so the effects in the multiplication layer which happen on a timescale of hundreds of picoseconds are not important. These effects are approximated by adding a Gaussian random variable to the time at which the photon arrives in the multiplication layer, which also account for the electronics timing resolution.

\begin{figure}[ht]
  \centering
  \includegraphics[scale=0.4]{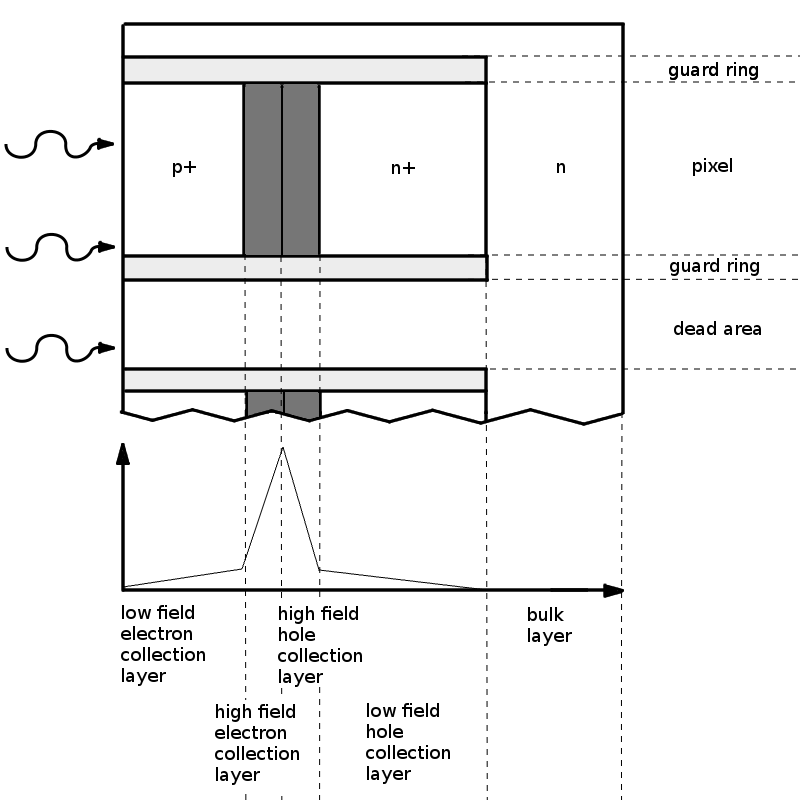}
  \caption{Inferred structure of the MPPC diode and corresponding electric field. The width and doping concentration of the p+, n+ implants is not known. Hence, the shape of the electric field is purely for illustration purpose.}
  \label{fig:mppcstructure}
\end{figure}

The device structure is another important consideration for the simulation.
The only details about the MPPC diode structure were disclosed by Hamamatsu photonics in~\cite{Yamamoto2007}.
A more detailed concept was proposed in~\cite{Oide2010} and explicitly described in~\cite{Orme2009}. This
concept was used with moderate consistency to discuss the photo-detection efficiency dependence with over-voltage at
4 different wavelengths, 470, 525, 590 and 625 nm.

Fig.~\ref{fig:mppcstructure} shows our conceptual representation of the one pixel
which was inspired by this earlier work.
The modeled structure of the MPPC is illustrated in Figure \ref{fig:mppcstructure}.
The electric fields in both the p+ and n+ regions are approximated by two
sections: a high field section with a large slope for the multiplication layer
and a low field section for the drift layer. The electric field is approximated
as being linear within each section for simplicity. The major parameters for
this model of the device are the maximum electric field in the low field regions
and the width of the low field regions. These values can be determined from
fitting to the measured data. The parameters in the high field sections do not
have a significant effect on the distribution of the tail due to the fact that a
potential avalanche occurs very quickly once a carrier enters those sections.
The avalanche probabilities must also be considered.  Electrons and holes do not
have the same probability of triggering an avalanche~\cite{Aull}. In the case of
the MPPC, electrons cause avalanches for carriers produced in the p+ layer while
holes cause them for carriers produced in the n+ and bulk layers. Only the ratio
between the hole and electron avalanche probabilities is important for the
distribution of the tail. The exact values of these probabilities will affect
the photon detection efficiency which is not important for the distribution of
the tail. For the purposes of the simulation the electron avalanche probability
was fixed at an estimated 0.8 and the hole avalanche probability was varied to
fit to the data. It is worth noting that this probability ratio and the depth of
the multiplication layer are effectively degenerate parameters. However, since
neither can be directly measured they must be fit together. The parameters that
need to be taken into consideration for the simulation are summarized in Table
\ref{tab:simulationparameters}.

\begin{table}
  \centering
  \begin{tabular}{p{7cm} p{2cm} p{3cm}}
    \toprule
    Parameter       & Source of value & Value \\
    \midrule
    Device          &             & T2K MPPC \\
    Temperature     &             & -60$^\circ$C \\
    Pixel pitch     & Datasheet   & 50$\mu$m \\
    Fill factor     & Datasheet   & 62\% \\
    Pixel count     & Datasheet   & 26 $\times$ 26 pixels \\
    Dark noise      & Measured    & $1 \times 10^{-5}$ normalized counts/ns \\
    Multiplication layer time spread & Fitted & 60 ps \\
    Low field electron collection layer width & Fitted & 3.0 $\mu$m \\
    High field electron collection layer width & Estimated & 0.2 $\mu$m \\
    High field hole collection layer width & Estimated & 0.2 $\mu$m \\
    Low field hole collection layer width & Fitted & 2.0 $\mu$m \\
    Bulk layer width & Estimated   & 300 $\mu$m \\
    Maximum electric field in low field electron collection layer & Fitted &
      0.08 V/$\mu$m \\
    Maximum electric field in low field hole collection layer & Fitted &
      0.50 V/$\mu$m \\
    Hole lifetime in bulk & Fitted & 300 ns \\
    Electron avalanche probability & Estimated & 0.8 \\
    Hole avalanche probability & Fitted & 0.27 \\
    Carrier diffusion constant  & Calculated  & \\
    Carrier mobility  & Calculated & \\
    Attenuation length  & Referenced & \cite{VirginiaSemiconductor} \\
    \bottomrule
  \end{tabular}
  \caption{Simulation parameters}
  \label{tab:simulationparameters}
\end{table}

\begin{figure}[ht]
  \begin{minipage}[b]{0.5\linewidth}
    \centering
    \includegraphics[scale=0.4]{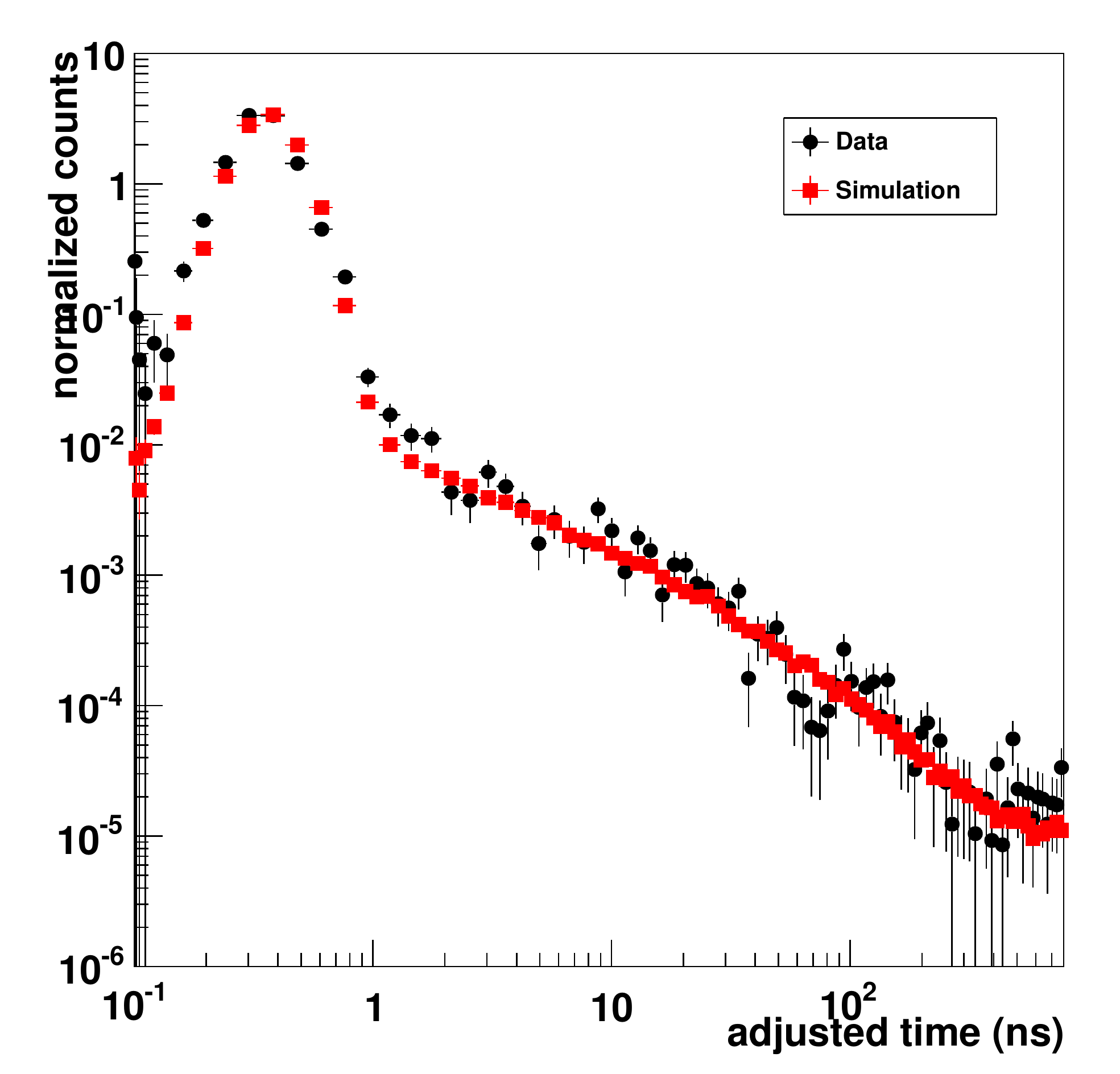}
    \caption{Timing distribution for collected and simulated data as a function
    of over-voltage for the Hamamatsu T2K MPPC using 637 nm light at
    -60$^\circ$C}
    \label{fig:t2ksim637}
  \end{minipage}
  \hspace{0.5cm}
  \begin{minipage}[b]{0.5\linewidth}
    \centering
    \includegraphics[scale=0.4]{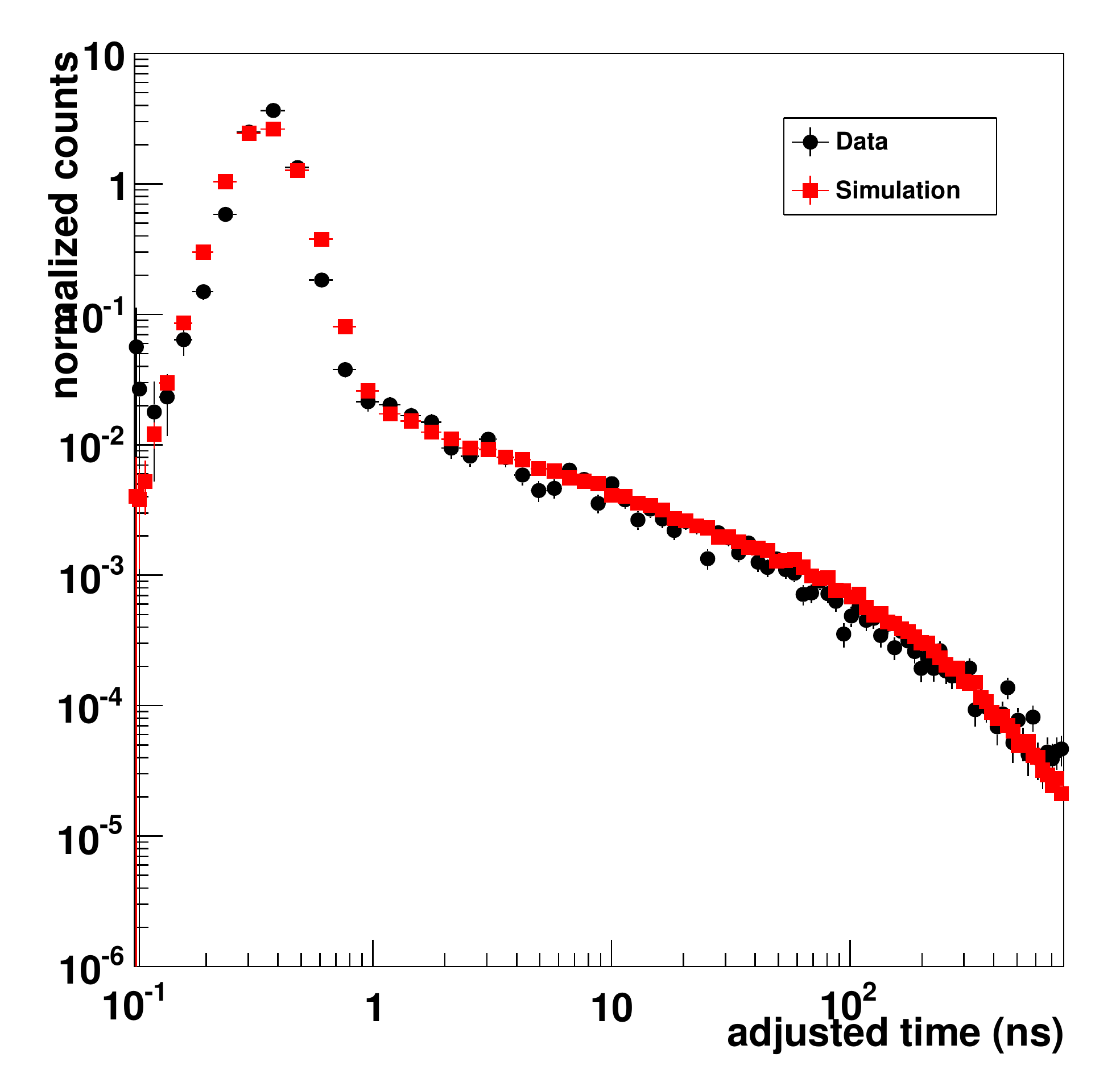}
    \caption{Timing distribution for collected and simulated data as a function
    of over-voltage for the Hamamatsu T2K MPPC using 820 nm light at
    -60$^\circ$C}
    \label{fig:t2ksim820}
  \end{minipage}
\end{figure}

With these parameters, the simulation produces results that agree well with the
data for all wavelengths over the full range of the distribution.  Comparisons
of both the data and the simulation are shown in Figures \ref{fig:t2ksim637} and
\ref{fig:t2ksim820} for 637 nm and 820 nm. For 405 nm and 467 nm light, the tail
is negligible as seen in the original data. These simulations show that this model of the MPPC is a very good representation
of the device. This has significant implications for after-pulsing. The photons
that are released during an avalanche have a wide spectra of wavelength and are
able to penetrate deep into the bulk. For those with longer wavelengths, the
probability that they produce a delayed avalanche is quite significant according
to this model. The same model can be used to simulate after-pulsing, replacing the external monochromatic beam by an internal source emitting isotropic light at the p-n junction with the wavelength spectrum measured in \cite{Mirzoyan2009}.

The results of the after-pulsing simulation are in qualitative agreement with T2K data suggesting that secondary photons from the electron avalanche are the
source of both the delayed pulses and after-pulse. By adding a buried layer that
would create an electric field pointing towards the substrate, it may be
possible to significantly reduce both the rate of after-pulse, and the delayed
avalanche tail. In 2013, HPK released a new device with much lower after-pulsing after we disclosed preliminary results. Investigating the new devices will certainly shed light on the source of after-pulsing for pre-2013 MPPCs.

\section{Performances of the new MPPCs}

The data were taken in the same way as before by operating the MPPC at -60$^\circ$C in order to measured delayed
avalanches over a $\mu$s time scale. The 637nm and 820nm laser heads were no longer available for this test and
they were replaced by a 777nm head for Hamamatsu PLP-10 that is qualitatively equivalent to 820nm,
the attenuation lengths being 10$\mu$ms at 777nm and 14 $\mu$m at 820nm. The 777nm timing distribution was not optimized for our controller leading to smearing of the prompt timing distribution without affecting the timing distribution of delayed avalanches. Because of these changes the data presented earlier cannot be compared directly with the new data.
Hence, new data were taken with a T2K MPPC for reference.

\begin{figure}[ht]
  \centering
  \includegraphics[scale=0.4]{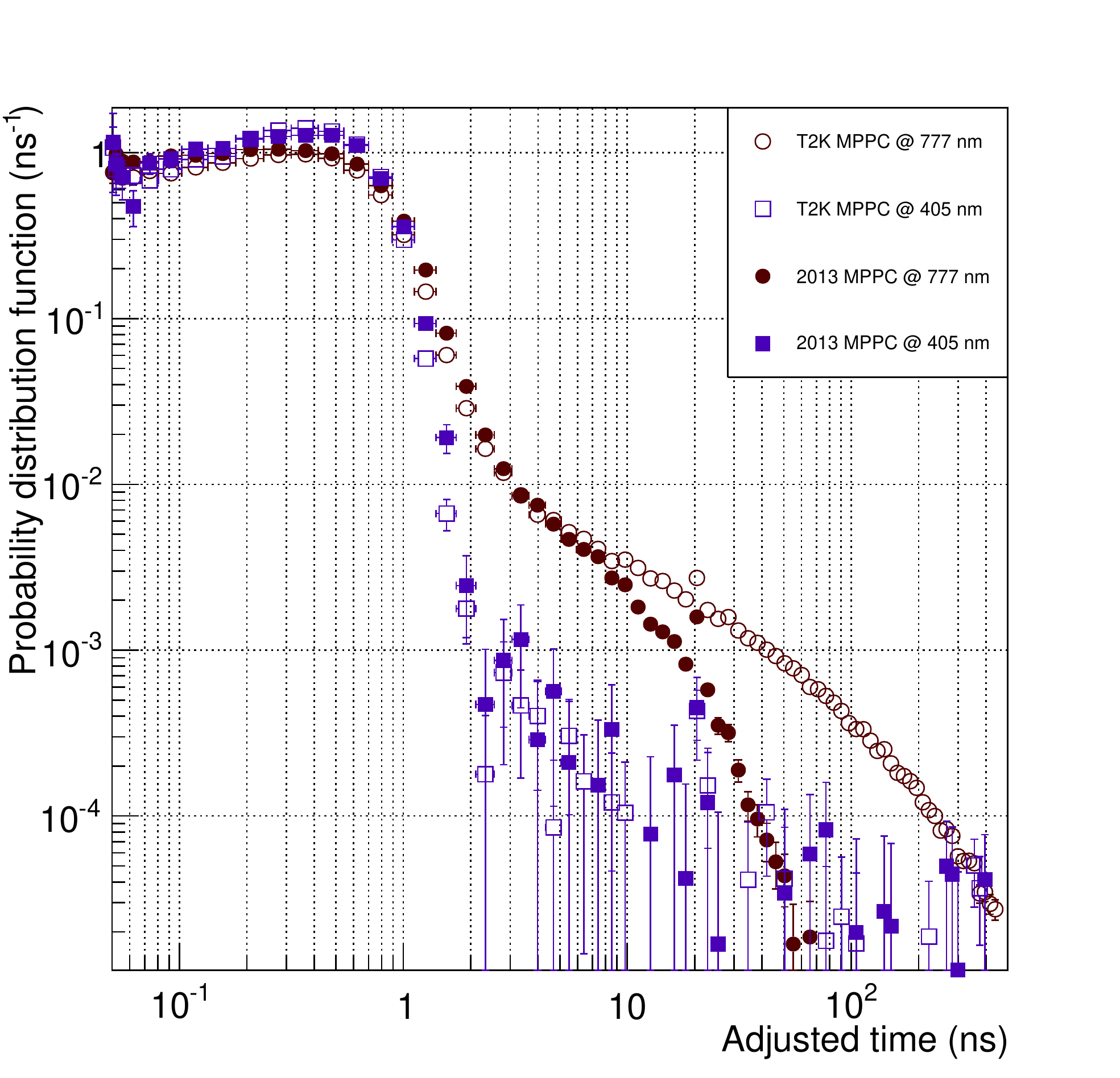}
  \caption{Comparison of delayed pulse probability distributions for old and
  new devices, normalized by probability of photon detection at
  -60$^\circ$C}
  \label{fig:ByArea}
\end{figure}

The timing distribution of the avalanches for the new and T2K MPPC are shown in Figure \ref{fig:ByArea}. The
distributions are essentially identical at 405nm showing no avalanches delayed by more than 2ns. On the
other hand, delayed avalanches are clearly visible at 777nm. The shape of the distribution for the new and T2K MPPCs
are very different however. The probability of delayed avalanches is significantly reduced for the new MPPC especially beyond
80ns. The first 3 rows of table\ref{tab:2013data} were obtained by integrating the timing distribution over different
timing windows for the 2 different wavelengths. The table shows that 1) the number of delayed avalanches with the
405nm laser is consistent with zero for both devices, 2) a reduction of the probability of delayed avalanches between
2 and 80ns by a factor of 2 for the new devices, 3) the probability of delayed avalanches after 80ns is consistent
with zero for the new device. Hence we infer that the new devices include a structure that either reduce the hole lifetime in the silicon substrate or prevent holes created
deep in the silicon bulk to diffuse back to the high field region. If infra-red photons are responsible for
after-pulse avalanches then a significant reduction of their rate is expected especially beyond 80ns. The situation
is somewhat unclear for avalanches occurring between 2 and 80ns because it is not possible to relate unambiguously
avalanche time with charge carrier creation point.

\begin{table}
  \centering
  \begin{tabular}{p{8.5cm} p{2.5cm} p{2.5cm}}
    \toprule
    Data                                                            &New MPPC         &T2K MPPC\\
    \midrule
    Avalanche probability in 2-500ns window @405nm                  &-0.7$\pm6.6\%$   &-0.04$\pm2.1\%$\\
    Avalanche probability in 2-80ns window @777nm                   &7.1$\pm0.2\%$    &13.5$\pm0.1\%$\\
    Avalanche probability in 80-500ns window @777nm                 &-0.3$1.1\%$      &4.5$\pm0.4\%$\\
    Dark noise rate at room temperature (kHz/mm$^2$)                &75.27$\pm0.73$     &315.9$\pm1.4$\\
    Average number of after-pulses per avalanche at -60$^\circ$C    &0.0066$\pm0.0002$  &0.1205$\pm0.0006$\\
    Fraction of after-pulse occurring after 80ns                    &-2.1$\pm1.6\%$   &35.5$\pm0.3\%$\\
    \bottomrule
  \end{tabular}
  \caption{Results of 2013 MPPC analysis at -60$^\circ$C at 1V overvoltage. The dark noise rate data was taken at room temperature (23-24.5$^\circ$C) at the recommended over-voltage of 3.06 and 1.24 overvoltage for the new and T2K MPPC respectively.}
  \label{tab:2013data}
\end{table}

Another expected consequence of preventing the diffusion of holes from the silicon bulk is reduction of the dark noise rate. Table\ref{tab:2013data} shows the dark noise rate measured at room temperature at the recommended operating voltage. The dark noise rate was measured by counting the fraction of events with no pulses within a 500ns window and applying Poisson statistics. The dark noise rate was normalized per unit area as the T2K MPPC has an active area of 1.44m $^2$ compare to 1 mm$^2$ for the new device. The dark noise rate is a factor of 4 lower for the new devices than for the T2K MPPC, which agrees with our expectation. However, the drop in dark noise rate may also be due to improved silicon purity.

\begin{figure}[ht]
  \centering
  \includegraphics[scale=0.4]{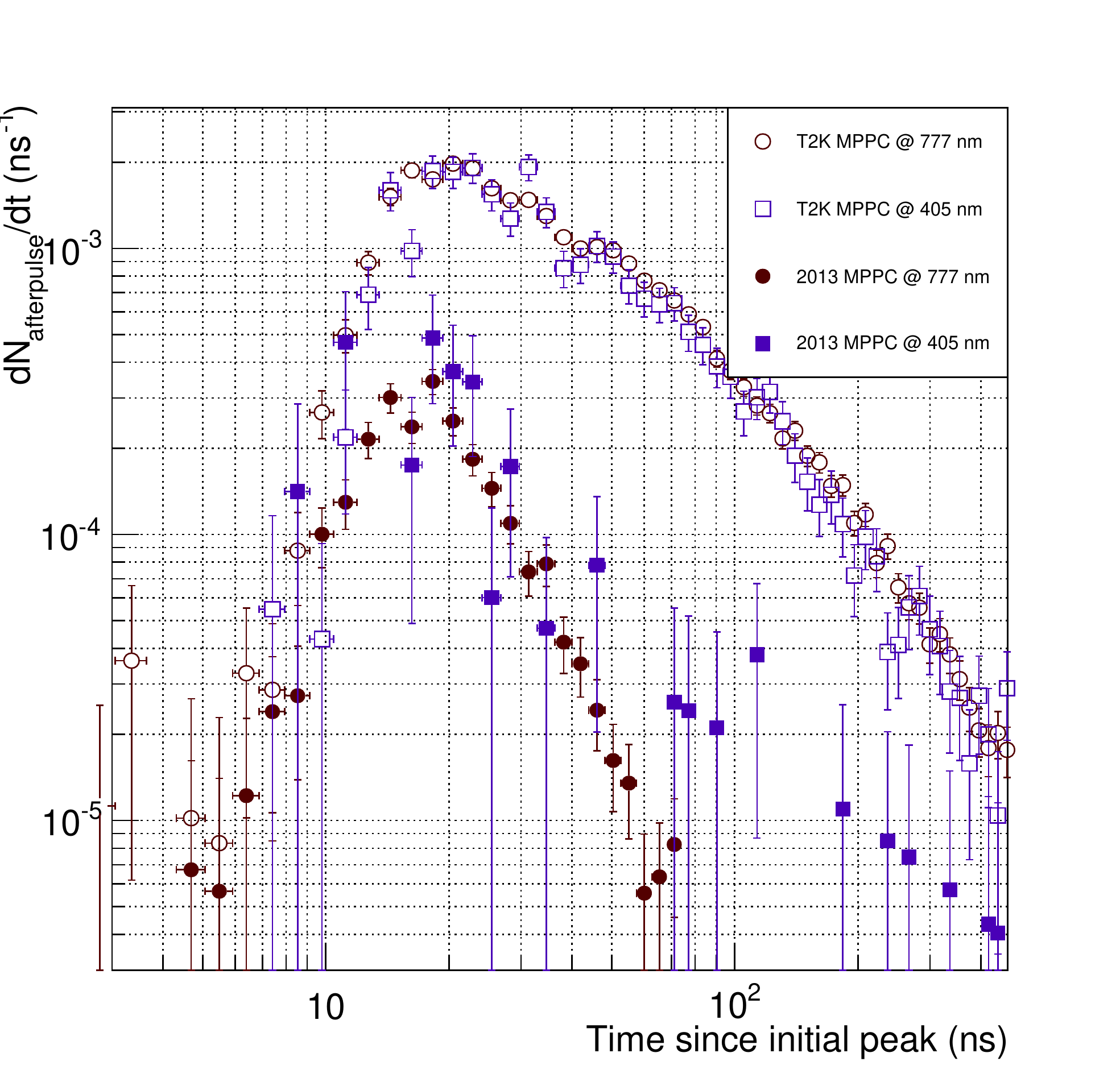}
  \caption{Comparison of afterpulse probabilty distributions for old and new
  devices at -60$^\circ$C}
  \label{fig:afterpulse2}
\end{figure}

After-pulsing was investigated from the same data that were used to measure the delayed avalanche timing distribution with a different event selection scheme. Events were selected such that one single pixel avalanche (i.e. no cross-talk) occurred in the 2ns wide prompt laser window with no earlier pulses (within 500ns). The timing distribution of the time difference between the trigger pulse (at time zero) and the pulses following was constructed. Events with more than one pulse following the trigger pulse are accepted. The distribution is normalized by the number of selected events, including events with no pulses following the trigger pulse. The integral of this distribution is then the number of after-pulse avalanches per parent avalanche, including possible after-pulse of after-pulse. The timing may be a bit skewed because of after-pulse of after-pulse but this distribution is easy to construct and dark noise can be easily subtracted out as a constant independent of time. The timing distribution of after-pulse avalanches is shown in Figure\ref{fig:afterpulse2} for the T2K and new MPPCs. It is independent of the laser wavelength. On the other hand, the new and T2K distribution are very different. The total average number of after-pulse is a factor 25 lower for the new MPPC as shown in table\ref{tab:2013data}. Furthermore, there are no after-pulse avalanches after 80ns with the new MPPC, which is consistent with the absence of delayed avalanches measured with the 777nm wavelength. Hence, we conclude that the source of after-pulse avalanches beyond 80ns for the pre-2013 MPPC was almost certainly due to infra-red photons. For avalanche occurring between 2 and 80ns after the parent, the situation is not as obvious because the probability of delayed avalanches occurring within 2 and 80ns at 777nm goes down by a factor of 2 while the after-pulsing rate between 2 and 80ns goes down by a factor of 10. Some of the difference can be explained by arguing that pixel recovery prevents the occurrence of after-pulse avalanches too close to their parents. The recovery time constant is indeed expected to be 13ns for 50$\mu$m pixels.

\section{Conclusions}

The timing distribution for MPPCs was measured using laser light between 405 nm and 820 nm. These distributions show that avalanches at the longer wavelengths can be delayed by hundreds of nanoseconds for pre-2013 devices. This dependence on wavelength cannot be explained by carrier trapping, and shows that holes produced in the silicon bulk can diffuse back to the avalanche region. A simulation was created which implemented this phenomenon, and it reproduces the data well. The simulations suggest that after-pulsing can be qualitatively explained by the same phenomenon changing the source of photons from external to internal and using the measured spectrum of photons emitted during avalanches. This work motivated Hamamatsu photonics to investigate a new device structure that was released in 2013. 

The timing distribution of the device introduced in 2013 shows a much shorter tail at longer wavelengths, as well as a significant reduction in the rate of after-pulsing when compared to the T2K MPPCs. The new devices show essentially zero probability for a 777nm photon to trigger an avalanches delayed by more than 80ns, while this probability was 4.5\% for T2K MPPCs.
The probability of an after-pulse occurring after 80ns also decreased from 35.5\% for T2K MPPCs to ~0\% for the new devices. The fact that both effects were eliminated after 80ns most likely implies that HPK changed the internal structure of the MPPC to prevent holes that are formed in the silicon bulk below the pixel by high wavelength photons from diffusing into the high field region. Overall after-pulsing was reduced by a factor 25, which is a very significant improvement allowing operation of the MPPC at higher overvoltages. A significant reduction in dark noise from 315.9kHz/$mm^2$ to 75.3kHz/$mm^2$ was also observed, which can be at least in part attributing to the reduction in the rate of avalanches generated by holes thermally generated in the silicon bulk and diffusing to the high field region. Hence, this work provides a compelling explanation to explain the improved MPPC performances starting in 2013 and it shows the power of probing such devices at several wavelengths, including infra-red wavelengths even though they are not relevant for most applications.

\section*{Acknowledgments}
We would like to thank Hamamatsu corporation for the loan of the 820nm laser head used in this study. 

\bibliographystyle{plain}

\end{document}